\documentclass{article}

\usepackage{arxiv}

\usepackage[utf8]{inputenc} 
\usepackage[T1]{fontenc}    
\usepackage{hyperref}       
\usepackage{url}            
\usepackage{booktabs}       
\usepackage{amsfonts}       
\usepackage{nicefrac}       
\usepackage{microtype}      
\usepackage{lipsum}		
\usepackage{graphicx}
\usepackage{natbib}
\usepackage{doi}

\usepackage{amsmath}
\usepackage{float}

\title{Estimating Nationwide High-Dosage Tutoring Expenditures: A Predictive Model Approach}

\date{August 9, 2024}	

\author{ \href{https://orcid.org/0000-0002-1977-9427}{\includegraphics[scale=0.06]{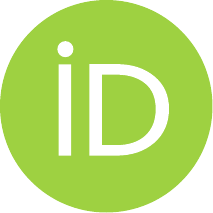}\hspace{1mm}Jason Godfrey}\\
	Strategic Data Fellow\\
	Harvard University\\
	Boston, MA 02138 \\
	\And
	\href{https://orcid.org/0000-0000-0000-0000}{\includegraphics[scale=0.06]{orcid.pdf}\hspace{1mm}Trisha Banerjee} \\
	Data Engineer\\
	Amazon Web Services\\
}

\renewcommand{\shorttitle}{\textit{arXiv} Template}

\hypersetup{
pdftitle={A template for the arxiv style},
pdfsubject={q-bio.NC, q-bio.QM},
pdfauthor={David S.~Hippocampus, Elias D.~Striatum},
pdfkeywords={First keyword, Second keyword, More},
}

\begin{document}
\maketitle

\begin{abstract}
    This study applies an optimized XGBoost regression model to estimate district-level expenditures on high-dosage tutoring from incomplete administrative data. The COVID-19 pandemic caused unprecedented learning loss, with K-12 students losing up to half a grade level in certain subjects. To address this, the federal government allocated \$190 billion in relief. We know from previous research that small-group tutoring, summer and after school programs, and increased support staff were all common expenditures for districts. We don't know how much was spent in each category. Using a custom scraped dataset of over 7,000 ESSER (Elementary and Secondary School Emergency Relief) plans, we model tutoring allocations as a function of district characteristics such as enrollment, total ESSER funding, urbanicity, and school count. Extending the trained model to districts that mention tutoring but omit cost information yields an estimated aggregate allocation of approximately \$2.2 billion. The model achieved an out-of-sample $R^2$=0.358, demonstrating moderate predictive accuracy given substantial reporting heterogeneity. Methodologically, this work illustrates how gradient-boosted decision trees can reconstruct large-scale fiscal patterns where structured data are sparse or missing. The framework generalizes to other domains where policy evaluation depends on recovering latent financial or behavioral variables from semi-structured text and sparse administrative sources.
\end{abstract}


\section{Introduction}

The COVID-19 pandemic has profoundly impacted the educational landscape, exacerbating existing inequalities and resulting in significant learning loss among K-12 students. This loss, quantified as up to half a grade level in some subjects, has prompted a vigorous response from educators and policymakers aiming to mitigate the long-term consequences of this disruption \citep{kuhfeld_projecting_2020, dorn_covid-19_2020}. In response, the federal government allocated approximately \$190 billion in relief funding through the Elementary and Secondary School Emergency Relief (ESSER) funds, a substantial investment aimed at addressing these educational deficits.

Among the various interventions proposed, high-dosage tutoring has emerged as one of the most promising strategies for mitigating learning loss \citep{nickow_impressive_2020}. High-dosage tutoring, characterized by intensive, small-group or one-on-one instruction, has been shown to consistently positively boost student achievement, particularly in reading and mathematics \citep{kraft_online_2022}. However, while the theoretical benefits of high-dosage tutoring are well-documented, less is known about the actual allocation of ESSER funds towards this intervention across different school districts.

This study seeks to bridge this knowledge gap by estimating district-level expenditures on high-dosage tutoring. Utilizing a custom dataset scraped from over 7,000 ESSER plans, we implement an optimized XGBoost regression model \citep{chen_xgboost_2016} to infer spending levels among districts that reference tutoring but do not report explicit dollar amounts. This approach treats financial inference as a predictive modeling problem—leveraging a broad set of district-level features such as urbanicity, total ESSER allocation, and school count—to impute unobserved values. Beyond its substantive findings, the paper contributes methodologically by demonstrating how ensemble learning can reconstruct incomplete administrative data at national scale, offering a transparent, reproducible framework for policy analytics.

\section{Literature Review}

The literature on educational interventions post-pandemic underscores the critical need for effective resource allocation \citep{dewey_federal_2024}. High-dosage tutoring, characterized by frequent, intensive sessions, has been widely recognized as a potent strategy for addressing learning loss, particularly for students from disadvantaged backgrounds \citep{nickow_impressive_2020}. Studies have consistently demonstrated the effectiveness of such tutoring in improving academic outcomes \citep{kraft_online_2022, robinson_accelerating_2021}.

Previous work emphasizes the profound impacts of the pandemic on student achievement and the necessity of targeted interventions to bridge the resultant gaps \citep{carbonari_impact_2024}. Research by \citet{gwynne_mitigating_2023, balfanz_increasing_2023-1} corroborates these findings, highlighting the efficacy of high-dosage tutoring in enhancing student performance in core subjects like reading and mathematics.

Recent meta-analyses reveal that small-group tutoring can significantly reduce achievement gaps, with the most substantial benefits observed in low-income student populations \citep{dietrichson_academic_2017}. These analyses show that interventions such as structured tutoring and feedback mechanisms are essential for improving educational outcomes among disadvantaged students. Additionally, studies indicate that the long-term economic benefits of improved educational outcomes far outweigh the initial costs of such programs, making high-dosage tutoring a cost-effective solution \citep{guryan_not_2023}.

Ensemble and supervised machine learning methods have increasingly been applied in educational, administrative, and policy-relevant settings to handle large, heterogeneous datasets and infer or predict outcomes where traditional econometric approaches may struggle. For example, a systematic review of predictive models in education finds that machine learning algorithms—including gradient boosting and tree-based methods—consistently outperform classical statistical models in forecasting student outcomes, managing non-linear relationships, and handling high-dimensional inputs \citep{almalawi_predictive_2024}. In one study, the tree-boosting algorithm XGBoost achieved superior performance in predicting learner performance on seven datasets compared with Item Response Theory models and standard logistic regression frameworks \citep{hakkal_xgboost_2024}. In policy- or finance-oriented contexts, gradient boosting methods (including XGBoost) have been shown to outperform alternatives in settings such as banking failure and financial distress modeling \citep{carmona_predicting_2019, lokanan_predicting_2024}. These findings signal the utility of ensemble learning for inferring latent or unreported values in administrative data, such as resource allocations across large populations—a methodology closely aligned with the present study’s objective of imputing district-level tutoring expenditures. Collectively, this line of research supports treating fiscal inference as a predictive modeling task using large-scale administrative inputs, offering novel methodological possibilities for education policy analytics.

\section{Data Sources}

\begin{figure}[H]
    \centering
    \includegraphics[width=\textwidth]{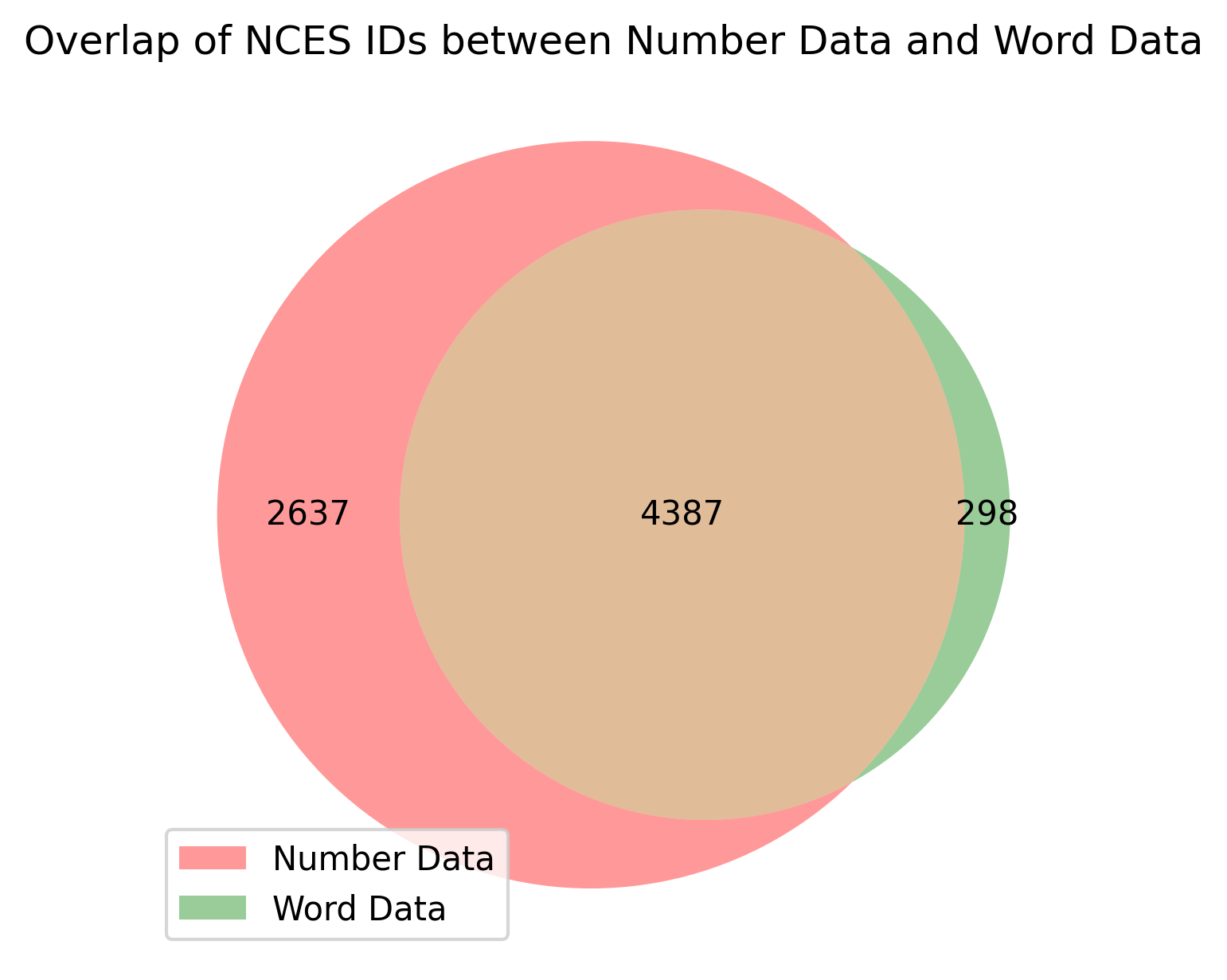}
    \caption{The overlap between the districts that have been manually annotated for budgetary items "Number Data," and the districts that have been passed through OCR and can be sorted by the presence/absence of tutoring "Word Data"}
    \label{fig:dataset-intersection}
\end{figure}

This study uses two custom data sources which both derive from the ESSER spending documents that each district was required to create in order to receive funding. The first dataset contains hand-scraped information from each of the ESSER documents regarding explicitly mentioned budget items such as tutoring, summer learning, and credit recovery. This dataset contains 7024 (roughly 36\% of total districts) unique districts with a combined enrollment of 41.1 million students, or roughly 83.45\% of estimated k12 enrollment in the USA. The second dataset contains 4685 district plans that have been passed through optical character recognition and their text can be searched. The intersection of these datasets is 4387 districts that can be analyzed for both the textual presence of "tutor*" and on specific expenditures that were manually extracted. This breakdown can be viewed in figure \autoref{fig:dataset-intersection}.

\begin{figure}[H]
    \centering
    \includegraphics[width=\textwidth]{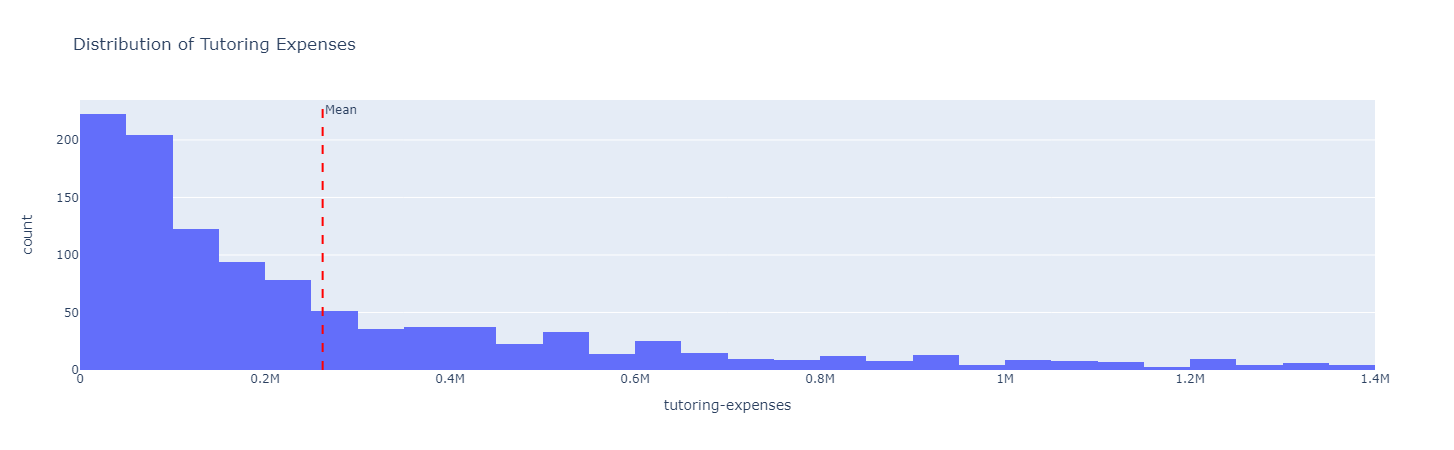}
    \caption{Distribution of tutoring expenses. Each bin is a \$50,000 increment.}
    \label{fig:expenses-histogram}
\end{figure}

Within the intersected dataset, 1232 districts defined how much ESSER money they allocated to tutoring. The mean tutoring allocation was \$263,213.68 with a standard deviation of \$296,770.74. This distribution can be seen in \autoref{fig:expenses-histogram}. Note that this histogram is drawn after removing outliers, such as Houston ISD that allocated 113.33 million to "High Dosage Tutorials."

This dataset, while incomplete, provides the best available insight into district-level expenditures of ESSER funds on all items not recorded by the federal government, such as high-dosage tutoring. All these plans are publicly available on their respective district websites and they contain no sensitive information.

\section{Methods}

In this study, we employed the XGBoost (Extreme Gradient Boosting) algorithm to predict tutoring expenses based on various features. XGBoost is an efficient and scalable implementation of gradient boosting, a powerful machine learning technique for regression and classification problems.

\subsection{Gradient Boosting Framework}

Gradient boosting builds an ensemble of weak learners, typically decision trees, by sequentially fitting new models to the residual errors made by previous models. The objective is to minimize the loss function $L(y_i, \hat{y}_i)$, where $y_i$ is the true value and $\hat{y}_i$ is the predicted value. Given a training dataset $\{(x_i, y_i)\}_{i=1}^n$, the model prediction $\hat{y}_i$ at iteration $t$ is updated as follows:

\begin{equation}
\hat{y}_i^{(t)} = \hat{y}_i^{(t-1)} + \eta f_t(x_i),
\end{equation}

where $\eta$ is the learning rate, and $f_t(x_i)$ is the weak learner fitted to the residual errors from the previous iteration. The objective function to be minimized is given by:

\begin{equation}
\mathcal{L}^{(t)} = \sum_{i=1}^n L(y_i, \hat{y}_i^{(t)}) + \sum_{k=1}^t \Omega(f_k),
\end{equation}

where $\Omega(f_k)$ is a regularization term that penalizes the complexity of the model, helping to prevent overfitting.

\subsection{XGBoost Algorithm}

XGBoost extends the gradient boosting framework with advanced features, including regularization, parallelization, and handling of missing values. The objective function in XGBoost can be written as:

\begin{equation}
\mathcal{L} = \sum_{i=1}^n l(y_i, \hat{y}_i^{(t-1)} + f_t(x_i)) + \Omega(f_t),
\end{equation}

where $l$ is a differentiable convex loss function that measures the difference between the prediction and the target. The regularization term $\Omega(f_t)$ is defined as:

\begin{equation}
\Omega(f_t) = \gamma T + \frac{1}{2} \lambda \sum_{j=1}^T w_j^2,
\end{equation}

where $T$ is the number of leaves in the tree, $w_j$ is the weight of leaf $j$, $\gamma$ is a regularization parameter for the number of leaves, and $\lambda$ is a regularization parameter for the leaf weights.

In estimating ESSER spending on tutoring, we utilized the trained XGBoost model to predict the expenses based on features such as school enrollment, number of schools, program allocations, and categorical variables for states and locales. The dataset was cleaned to remove any rows with missing or zero values for tutoring expenses, and outliers were filtered using the Interquartile Range (IQR) method. Categorical variables were then encoded as dummy variables. After training the XGBoost model, we calculated the residuals, defined as the difference between the actual values \(y\) and the predicted values \(\hat{y}\). The standard deviation of these residuals, \(\sigma_{\text{res}}\), was used to establish an error margin for the predictions. For each prediction, the low estimate was calculated as \(\hat{y} - \sigma_{\text{res}}\) and the high estimate as \(\hat{y} + \sigma_{\text{res}}\), providing a range that accounts for the variability in the model’s predictions.

\subsection{Model Training and Evaluation}

We utilized the XGBoost implementation from the \texttt{xgboost} Python library. The following steps outline the process:

\begin{enumerate}
    \item \textbf{Data Preparation}: The dataset was cleaned by removing rows with missing or zero values for tutoring expenses. Categorical features were encoded using one-hot encoding.
    \item \textbf{Train-Test Split}: The data was split into training and test sets using an 80-20 split.
    \item \textbf{Model Training}: The XGBoost model was trained on the training set using 100 estimators. The hyperparameters were chosen based on standard recommendations and adjusted through cross-validation.
    \item \textbf{Model Evaluation}: The model's performance was evaluated on the test set using metrics such as Mean Absolute Error (MAE), Mean Squared Error (MSE), Root Mean Squared Error (RMSE), R² Score, Adjusted R² Score, and Mean Absolute Percentage Error (MAPE).
\end{enumerate}

We employed an Optuna-based hyperparameter search within an XGBoost regression framework to estimate the share of ESSER funds allocated to high-dose tutoring (HDT). In this approach, we defined a scalar-valued objective function that performs five-fold cross-validation on the training set using the \texttt{XGBRegressor} estimator. Specifically, the objective function receives a set of hyperparameters from the trial, instantiates an \texttt{XGBRegressor} with those hyperparameters, and returns the negative mean of the cross-validated root mean squared errors (RMSE). The search space encompassed nine parameters: \texttt{n\_estimators} (integer, 50--1000), \texttt{max\_depth} (integer, 3--15), \texttt{learning\_rate} (float, $10^{-4}$--0.5, log-scale), \texttt{subsample} (float, 0.5--1.0), \texttt{colsample\_bytree} (float, 0.5--1.0), \texttt{min\_child\_weight} (integer, 1--7), \texttt{reg\_alpha} (float, $10^{-8}$--1, log-scale), \texttt{reg\_lambda} (float, $10^{-8}$--1, log-scale), and \texttt{gamma} (float, $10^{-8}$--1, log-scale). The hyperparameters were explored via Tree-structured Parzen Estimator (TPE), which adaptively models the density of favorable versus unfavorable hyperparameter regions. Each trial yielded a distinct hyperparameter combination, and the scoring function reported the mean of five-fold cross-validated RMSE values, ensuring robust estimation of generalization error. The study was executed for 250 trials, after which the optimized hyperparameter set was extracted via \texttt{study.best\_params}. Lastly, we serialized the values of the objective function across all trials for subsequent inspection and reproducibility.

\section{Results}

The results are organized into three components. First, we assess the model’s internal fit and out-of-sample performance during training and validation. Second, we apply the trained model to the broader set of districts that mention tutoring but do not report expenditure amounts, in order to generate predicted spending estimates. Third, we analyze residuals from the fitted model to quantify uncertainty and construct an estimated range for total district-level allocations toward high-dosage tutoring.

\subsection{Model Fitting and Training}

Model fitting refers to the process by which the XGBoost algorithm iteratively minimizes prediction error on the training data by adjusting its internal parameters to capture statistical relationships between explanatory variables and the response. In this study, the model was trained on districts that explicitly reported tutoring expenditures, using features such as total ESSER allocation, student enrollment, number of schools, and urbanicity. The training process involved repeated gradient-based optimization to reduce the residual difference between predicted and observed expenditures, while the hyperparameter search ensured an optimal balance between model flexibility and generalization.

The hyperparameter optimization procedure, conducted through 250 Optuna trials, converged on a configuration characterized by moderate depth and conservative regularization. The final model used 457 estimators (\texttt{n\_estimators}=457), a tree depth of 3 (\texttt{max\_depth}=3), and a learning rate of 0.0108, promoting incremental updates that reduce the risk of overfitting. Subsampling rates for rows and features (\texttt{subsample}=0.7906; \texttt{colsample\_bytree}=0.8462) introduced stochasticity to stabilize the model, while minimal regularization parameters (\texttt{reg\_alpha}, \texttt{reg\_lambda}, and \texttt{gamma} near $10^{-7}$) indicated that the model achieved satisfactory performance without requiring heavy penalization of complexity.  

The fitted model explained approximately 35.8\% of the variance in reported tutoring expenditures ($R^2 = 0.3581$), with an adjusted $R^2 = 0.3361$ accounting for model complexity. While a portion of variance remains unexplained—reflecting the heterogeneity of district reporting and local fiscal practices—these results indicate that the model effectively captures systematic relationships among key district-level predictors of tutoring expenditure.

\subsection{Application to Unreported Cases}

Following training, the fitted model was applied to districts that referenced tutoring in their ESSER plans but did not disclose a dollar amount. In this application phase, the model’s parameters were held constant; no additional fitting or optimization was performed. Instead, the model produced predicted values based on the same input variables that were found to be predictive in the training data.  

This application yielded an estimated total of approximately \$2.2 billion in district-level allocations for high-dosage tutoring. While this figure does not represent a definitive accounting total, it provides an empirically grounded estimate derived from observed data and model-inferred relationships. Given the wide range of anecdotal estimates in circulation, spanning from roughly \$700 million to \$7.5 billion, this result offers a data-driven midpoint anchored in reproducible statistical inference.

\subsection{Residual Analysis and Uncertainty Estimation}

\begin{figure}[H]
    \centering
    \includegraphics[width=\textwidth]{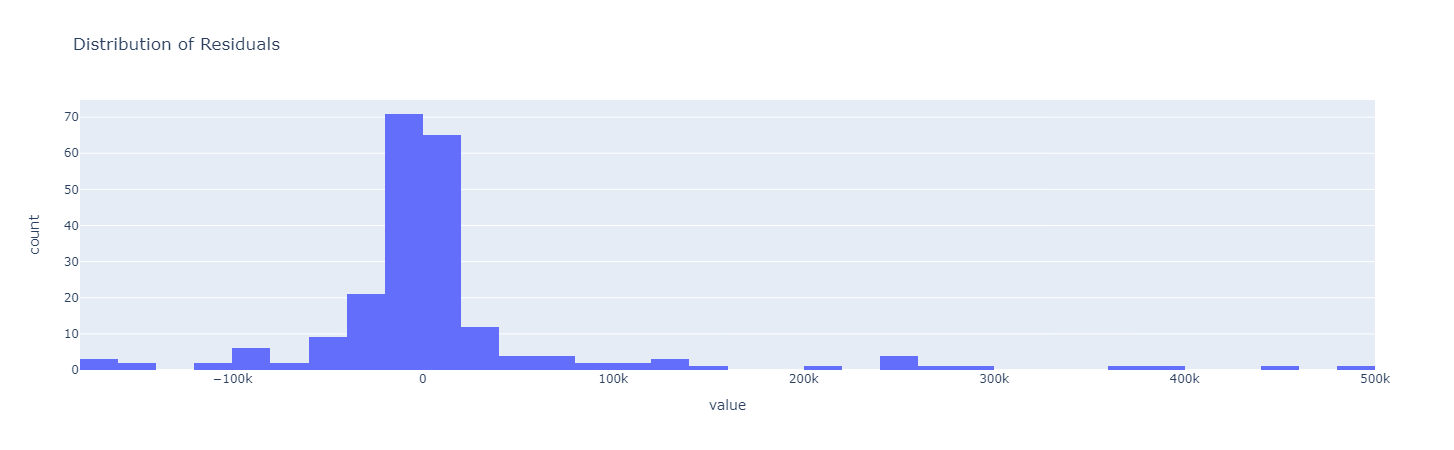}
    \caption{Distribution of residuals. Each bin is a \$20,000 increment.}
    \label{fig:residuals-histogram}
\end{figure}

To evaluate the reliability of these estimates, we examined the distribution of residuals from the fitted model. In other words, we inspect the differences between observed and predicted tutoring expenditures among districts with known values. As shown in \autoref{fig:residuals-histogram}, the residuals exhibit a narrow, symmetric distribution centered near zero, with most deviations falling between \$0 and \$20{,}000. This distribution suggests that the model’s predictive errors are both limited in magnitude and unbiased in direction.

While the majority of residuals cluster tightly around zero, several pronounced outliers are visible in the upper tail of the distribution. These high-magnitude residuals indicate a small number of districts where the model substantially underestimates actual expenditures. The skew toward positive residuals suggests that, in these instances, reported spending exceeded the level predicted by the model—potentially reflecting unique local budget priorities, multi-year tutoring contracts, or reporting anomalies. Although infrequent, these outliers underscore the heterogeneity of district-level spending behavior and warrant consideration when interpreting aggregate estimates.

We used the standard deviation of residuals ($\sigma_{\text{res}}$) as a measure of typical prediction uncertainty. By adding and subtracting this value from each model prediction, we derived upper and lower bounds for total predicted expenditures. This approach yields an estimated range of \$1.87 billion to \$3.85 billion in total district-level spending on high-dosage tutoring. The width of this range reflects uncertainty stemming primarily from data incompleteness and the inherent heterogeneity of district plans.

\subsection{Interpretation}

Taken together, these analyses reveal consistent evidence that districts prioritized high-dosage tutoring as a central strategy for academic recovery. Even under the most conservative assumptions, estimated allocations approach \$2 billion nationwide; at the higher end, they exceed \$3.8 billion. This finding positions high-dosage tutoring as a substantively and fiscally significant category of ESSER-funded interventions.

More, the modeling approach demonstrates that machine learning techniques can reconstruct credible expenditure patterns from incomplete administrative data. The model’s performance, coupled with the transparency of its validation, provides a replicable framework for analyzing large-scale education finance data where formal reporting mechanisms are incomplete or inconsistent. In doing so, it bridges an important methodological gap between policy analysis, inferential analysis, and machine learning in education research.

\section{Scholarly Significance}

This study advances our understanding of how federal recovery funds have been deployed to combat pandemic-related learning loss, offering the first systematic, data-driven estimate of district-level investments in high-dosage tutoring. Using a tuned XGBoost regression model on a uniquely constructed dataset of over 8,000 ESSER plans, we demonstrate how machine learning methods can illuminate opaque patterns of educational spending that elude conventional reporting systems. The model’s explanatory power and predictive precision provide an empirically grounded estimate of roughly \$2.2 billion in ESSER allocations to HDT, a figure that, while bounded by uncertainty, anchors the national conversation in evidence rather than conjecture.

The implications are twofold. First, this analysis demonstrates that a substantial share of relief funding was directed toward one of the most empirically validated interventions available. Methodologically, this work contributes to the applied machine learning literature by demonstrating how gradient boosting can be used for fiscal inference in understructured administrative datasets. The framework generalizes to other educational or policy contexts where missingness and heterogeneity hinder econometric analysis. Together, these contributions suggest that even in fragmented policy environments, statistical learning techniques can restore a degree of transparency and accountability to public education finance.

Yet, the findings also expose the limits of current data infrastructure. The wide confidence range surrounding estimated tutoring expenditures underscores a fundamental problem: despite historic federal investment, no unified reporting system exists to track how funds were actually used. Without standardized, machine-readable documentation of district spending, future researchers and policymakers alike will continue to operate in partial darkness.

In sum, this work demonstrates both what is possible and what remains undone. By applying advanced analytic methods to an incomplete but vital record, we show how high-dosage tutoring has become a cornerstone of pandemic recovery efforts—and how much more could be learned, and accomplished, with transparent, interoperable data systems. The path forward lies in ensuring that the evidence we already have can be seen, shared, and acted upon.

\bibliographystyle{unsrtnat}
\bibliography{references}  






\newpage
\appendix
\section{Figures and Tables}

\begin{table}[t]
\centering
\begin{tabular}{lcc}
\toprule
\textbf{Parameter} & \textbf{Min} & \textbf{Max} \\
\midrule
n\_estimators     & 51           & 948          \\
max\_depth        & 3            & 15           \\
learning\_rate    & 1.06e-4      & 0.327        \\
subsample         & 0.50         & 0.986        \\
colsample\_bytree & 0.50         & 0.991        \\
min\_child\_weight & 1          & 7            \\
reg\_alpha        & 1.01e-8      & 0.919        \\
reg\_lambda       & 1.34e-8      & 0.796        \\
gamma             & 1.31e-8      & 0.645        \\
\bottomrule
\end{tabular}
\caption{Overall parameter search space and observed ranges. A total of 150 iterations were conducted.}
\end{table}

\begin{table}[t]
\centering
\begin{tabular}{cccccc}
\toprule
\textbf{Iter.} & \textbf{Value} & \textbf{n\_estim.} & \textbf{max\_depth} & \textbf{learn\_rate} & \textbf{min\_child\_w.}\\
\midrule
15 & 1.601757e+6 & 51  & 6  & 0.0178   & 3 \\
42 & 2.141770e+6 & 465 & 9  & 0.0806   & 4 \\
64 & 1.801879e+6 & 708 & 3  & 1.07e-4  & 4 \\
71 & 1.605657e+6 & 342 & 6  & 0.00072  & 4 \\
99 & 2.121397e+6 & 775 & 15 & 0.00046  & 7 \\
\bottomrule
\end{tabular}
\caption{Five illustrative iterations. We include one near the best (lowest) value, one near the worst (highest) value, and others that highlight extremes in hyperparameters such as very low learning\_rate or large n\_estimators.}
\end{table}

\vspace{1ex}
\textit{Note:} These iterations were chosen to cover the best and worst performance, plus cases of extreme parameter settings that illustrate the diversity of the search.


\end{document}